\documentclass[12pt]{iopart}

\usepackage{amssymb,latexsym}
\usepackage{graphicx}
\usepackage{graphics}
\usepackage{setstack}
\usepackage{bm}
\usepackage{color}

\usepackage{iopams}  
\usepackage[colorlinks,bookmarks=true,citecolor=blue,linkcolor=red,urlcolor=blue]{hyperref}

\newcommand{\masa}[1]{\textcolor{black}{#1}}

\begin{document}

\title{Correlations and entanglement in flat band models with variable Chern numbers}

\author{Masafumi Udagawa}

\address{Department of Applied Physics, University of Tokyo, Hongo 7-3-1, Bunkyo-ku, Tokyo 113-8656}
\ead{udagawa@ap.t.u-tokyo.ac.jp}

\author{Emil J. Bergholtz}

\address{Dahlem Center for Complex Quantum Systems, Freie Universit\"at Berlin, Arnimallee 14, 14195 Berlin, Germany}
\ead{ejb@physik.fu-berlin.de}

\begin{abstract}
We discuss a number of illuminating results for tight binding models supporting a band with variable Chern number, and illustrate them explicitly for a simple class of two-banded models. First, for models with a fixed number of bands, we show that the minimal hopping range needed to achieve a given Chern number $C$ is increasing with $C$, and that the band flattening requires an exponential tail of long-range processes. We further verify that  the entanglement spectrum corresponding to a real-space partitioning contains $C$ chiral modes and thereby complies with the archetypal correspondence between the bulk entanglement and the edge energetics. Finally, we address the issue of interactions and study the problem of two interacting particles projected to the flattened band as a function of the Chern number. Our results provide valuable insights for the full interacting problem of a partially filled Chern band at variable filling fractions and Chern numbers. 
\end{abstract}
\maketitle

\section{Introduction}
The discovery \cite{tsui} and basic understanding \cite{Laughlin83} of the fractional quantum Hall effect (FQHE) have fundamentally transformed our understanding of the macroscopic states of matter.
The FQHE provides an example of a many-body state that does not fit in the conventional Ginzburg-Landau's paradigm, and urged us to
introduce the concept of topological quantum order to characterize this novel state of matter and it stands out as the only experimentally accessible system where the non-trivial quantum order is realized  \cite{wenbook}. Notably, the experimental stage of FQHE is semiconductor interfaces, where the magnetic length of the system well surpasses the microscopic lattice scale. Due to the virtue of this separation of length scales, the universal description of FQHE is available in terms of field theory, without suffering the influence from the short wave-length degrees of freedom. While this idealized setting has greatly facilitated the theoretical developments, it also underscores a major reason why the FQHE has not been seen practical applications e.g., in the context of topological quantum computing \cite{topocomp}, namely that the low density implies that the effective energy scale set by interactions is miniscule.

Recently, there has been great renewed interest alternative platforms for FQHE physics driven by the motivation to promote interaction scale to facilitate applications alongside with the insight that they uncover new and unexpected physics qualitatively beyond the standard continuum setting (see Refs.~\cite{Emilreview,otherreview} for recent reviews). 
This interest can be traced back to a much earlier work by Haldane who realized that in (non-interacting) systems with broken time-reversal symmetry, the electron bands can acquire 
an analogous property to the Landau level structure and exhibits quantized Hall conductivity even when the effective magnetic field averages to zero \cite{haldanemodel}. This class of electronic states is called (integer) Chern insulators, and have been theoretically predicted in a broad range of systems, and very recently also experimentally realized \cite{Chernexp}. 
Even more strikingly, the recent theoretical insight is that electron interaction are capable of transforming the Chern insulator into the non-trivial many-body state with topological quantum order. Indeed, it was pointed out that FQHE takes place in the partially filled Chern band, if the energy dispersion is sufficiently flat, compared with the typical scale of electron interaction \cite{kapit,chernins1,chernins2,chernins3,cherninsnum1,cherninsnum2}.

The transformation of the Chern insulators into the fractional quantum Hall state, generally called fractional Chern insulators (FCIs), opens a rich and fertile research area which is presently experiencing a rapid development \cite{Emilreview,otherreview}.
The Chern insulators are ubiquitous in a wide range of the solid state \cite{digital, ifw2, max, c2, grapheneFCI, organometallic,Martin,Akagi1,Akagi2} as well as cold atom \cite{DipolarTFB2, opticalflux,coopermoessner} systems, including the system with magnetic flux \cite{haldanemodel}, the magnetically ordered system with non-coplanar magnetic moments \cite{Martin,Akagi1,Akagi2},
and the orbital-ordered system with broken time-reversal symmetry \cite{ifw2}. FCIs have even been suggested to form as collective states of light (photons) \cite{light}. The variety of the mother systems promises a high controllability of the Chern insulators.
Among the rich tuning parameters existing in these Chern insulators, it is interesting to look at the Chern number $C$.
In contrast to the Landau level in the continuous system, where $C=1$ is always realized, it is possible to assign variable
integer to each band in the Chern insulator \cite{c2,max,dassarma,paramekanti}. 

In fact, the controllable Chern number gives a rich variety to the FCIs qualitatively distinct from conventional fractional quantum Hall states once short range repulsion is taken into account \cite{ChernN,ChernTwo,ChernN2,Grushin,disloc,TTFCI,layla,yangle,latticepp,sunnew}. Despite impressive progress in a number of special cases, the flat band models with $|C|>1$ are much less studied and the understanding of the interacting problem at generic filling fractions is far less advanced than in the $|C|=1$ bands, by virtue of the fruitful Landau level analogy applicable in the former case. 

In this paper, we take a step back and focus on some of the basic properties of Chern band models with variable $C$. For this purpose, we devise a  recipe for constructing convenient tight-binding models composed of two degrees of freedom at each site,
which leads to an electronic state composed of two bands, to which arbitrary integer Chern number can be assigned (depending on the tight binding parameters). The energy dispersion of this model can be made systematically flat, by introducing long-range transfer integrals and we study the effect of truncating these long range terms. In particular, we do this by computing a number of basic properties of this model, such as the ratio of band flatness to the energy gap, the Berry curvature, and one-body entanglement spectrum \cite{Peschel,Peschel2}. Albeit being of most interest for interacting systems \cite{LiH}, the entanglement spectrum is also known to provide useful information about topological phases free-fermions \cite{Turner2010,Fidkowski,Turner2012,Hughes, Sondhi,Budich,maria}, which we verify in the present case by recovering the bulk-edge correspondence between edge spectra and bulk entanglement. 
We further consider the two-particle interaction problem projected to the bands with variable $C$. As known in the context of the FQHE from Haldane's pseudo-potential analysis \cite{haldane83}, key insights into the full many-body problem can be sometimes gained from the spectrum of two-body problem. In the context of $C=1$ FCIs this concept has recently proven useful despite the fact that full rotational symmetry is lacking in the system \cite{andreas}. 

The organization of the paper is as follows. In Section \ref{model}, we describe the recipe for the two-band models
with arbitrary Chern number and in Section \ref{finiterange} we investigate the effect of truncating the range of the hopping in flattened models. In Section \ref{entanglement_formulation}, we analyze one-body entanglement spectrum and in Section \ref{twobodyproblem} we study the interacting  two-body problem projected to a Chern band.  Finally in Section \ref{discussion}, we briefly discuss the relevance of our results, connect our findings for a specific model to generic Chern band models and relate them to pertinent recent literature.

\section{Setup and basic properties}
In this Section, we describe the formulation of a simple two band model which we will use as a workhorse for illustrating the general points we want to make. We also provide a number of instructive examples and in particular discuss the effects of a truncated hopping range.
\subsection{An instructive two band model}
\label{model}
We start with the Hamiltonian
\begin{eqnarray}
\mathcal{H} = \sum_{\mathbf k}c^{\dag}_{{\mathbf k}\alpha}\mathcal{H}({\mathbf k})_{\alpha\beta}c_{{\mathbf k}\beta},
\label{Hamiltonian}
\end{eqnarray}
where
\begin{eqnarray}
\mathcal{H}({\mathbf k}) = (\sin k_x)\sigma_x + (\sin k_y)\sigma_y + (m + \cos k_x + \cos k_y)\sigma_z
\label{HofK}
\end{eqnarray}
is defined on a square lattice \masa{with $N_x\times N_y$ sites} and the Pauli matrices, $\sigma_i$, define an internal degree of freedom at each lattice point \cite{XiaoLiangQi2008}. The band structure of this model is composed of two bands, and the Chern number $C$ of the
lower band can be classified \cite{XiaoLiangQi2006}, according to the value of $m$, as
\begin{eqnarray}
C=
\left\{\begin{array}{ll}
1 & {\rm for}\ 0 < m < 2\\
-1 & {\rm for}\ -2 < m < 0\\
0 & {\rm otherwise}
\end{array}\right.
\label{Qi_C}
\end{eqnarray}
We set $m=1$ throughout this paper, which will ensure that we stick to the topologically non-trivial regime. By introducing ${\mathbf d}$-vector, 
\begin{eqnarray}
\left\{\begin{array}{l}
d_x({\mathbf k}) = \sin k_x\\
d_y({\mathbf k}) = \sin k_y\\
d_z({\mathbf k}) = m + \cos k_x + \cos k_y
\label{Qi_d}
\end{array}\right.
\end{eqnarray}
the Bloch Hamiltonian (\ref{HofK}) can be conveniently written as
\begin{eqnarray}
\mathcal{H}({\mathbf k}) = {\mathbf d}({\mathbf k})\cdot{\bm\sigma} . 
\label{Hofd}
\end{eqnarray}

The ${\mathbf d}$ vector representation directly leads to a geometrical expression for the Chern number:
\begin{eqnarray}
C = \frac{1}{4\pi}\int\limits\ dk_x\int\limits\ dk_y\ \hat{\mathbf d}\cdot\Bigl(\frac{\partial \hat{\mathbf d}}{\partial k_x}\times\frac{\partial \hat{\mathbf d}}{\partial k_y}\Bigr)\ ,
\label{2bandChern}
\end{eqnarray}
where $\displaystyle\hat{\mathbf d}\equiv{\mathbf d}({\mathbf k})/|{\mathbf d}({\mathbf k})|$ is the unit length vector parallel to ${\mathbf d}({\mathbf k})$. By regarding $\hat{\mathbf d}$ as the mapping from Brioullion zone (BZ) to the sphere surface $\displaystyle \hat{\mathbf d}: [0, 2\pi)\times[0, 2\pi)\rightarrow S^2$, the Chern number $C$ acquires the geometrical meaning: $C$ represents the wrapping number of this mapping. According to (\ref{Qi_C}), the ${\mathbf d}$-vector chosen as (\ref{Qi_d}) wraps the sphere only one time for $0<m<2$, while ${\mathbf k}=(k_x, k_y)$ sweeps the BZ. The integrand $B({\mathbf k})\equiv \hat{\mathbf d}\cdot\Bigl(\frac{\partial \hat{\mathbf d}}{\partial k_x}\times\frac{\partial \hat{\mathbf d}}{\partial k_y}\Bigr)$ is the Berry curvature which can be interpreted as a magnetic field in reciprocal space. Physically, the Chern number corresponds to the number of current carrying chiral edge states, which directly gives the quantized transverse conductivity $\sigma_{xy}=C\frac{e^2}{h}$ for a filled, hence bulk insulating, band carrying Chern number $C$ \cite{tknn,avron,qniu}.

In order to generate a two-band model with arbitrary Chern number from Hamiltonian (\ref{Hofd}), 
let us first introduce polar coordinate and express ${\mathbf d}$-vector (\ref{Qi_d}) as
\begin{eqnarray}
\Bigr(d_x(k), d_y(k), d_z(k)\Bigr)
= |{\mathbf d}({\mathbf k})|\Bigr(\sin\theta_k\cos\phi_k, \sin\theta_k\sin\phi_k, \cos\theta_k\Bigr)
\end{eqnarray}
A key observation is that the simple replacement 
\begin{eqnarray}
\phi_k
\rightarrow N \phi_k
\end{eqnarray}
leads to the $N$-fold wrapping of the mapping $[0, 2\pi)\times[0, 2\pi)\rightarrow S^2$. Hence, by defining a new ${\mathbf d}$-vector
\begin{eqnarray}
{\mathbf d}^{(N)}({\mathbf k})&\equiv &
\Bigr(d^{(N)}_x({\mathbf k}), d^{(N)}_y({\mathbf k}), d^{(N)}_z({\mathbf k})\Bigr) =\nonumber\\ &=&\Bigr(\sin\theta_k\cos(N\phi_k), \sin\theta_k\sin(N\phi_k), \cos\theta_k\Bigr) \ .
\end{eqnarray}
we arrive at the Hamiltonian
\begin{eqnarray}
\mathcal{H}^{(N)}\equiv\sum\limits_{\mathbf k}c^{\dag}_{{\mathbf k}\alpha}\hat{\mathbf d}^{(N)}({\mathbf k})\cdot{\bm\sigma}_{\alpha\beta}c_{{\mathbf k}\beta},
\label{Ham_dn}
\end{eqnarray}
which has a lower band with Chern number $N$ (the upper band has $C=-N$). Furthermore, since the Hamiltonian of the form (\ref{Ham_dn}) has eigenvalues $\pm|\hat{\mathbf d}^{(N)}({\mathbf k})|$,
this system has completely flat bands at $\varepsilon({\mathbf k})=\pm 1$.

\begin{figure}[t]
\begin{center}
\includegraphics[width=0.79\textwidth]{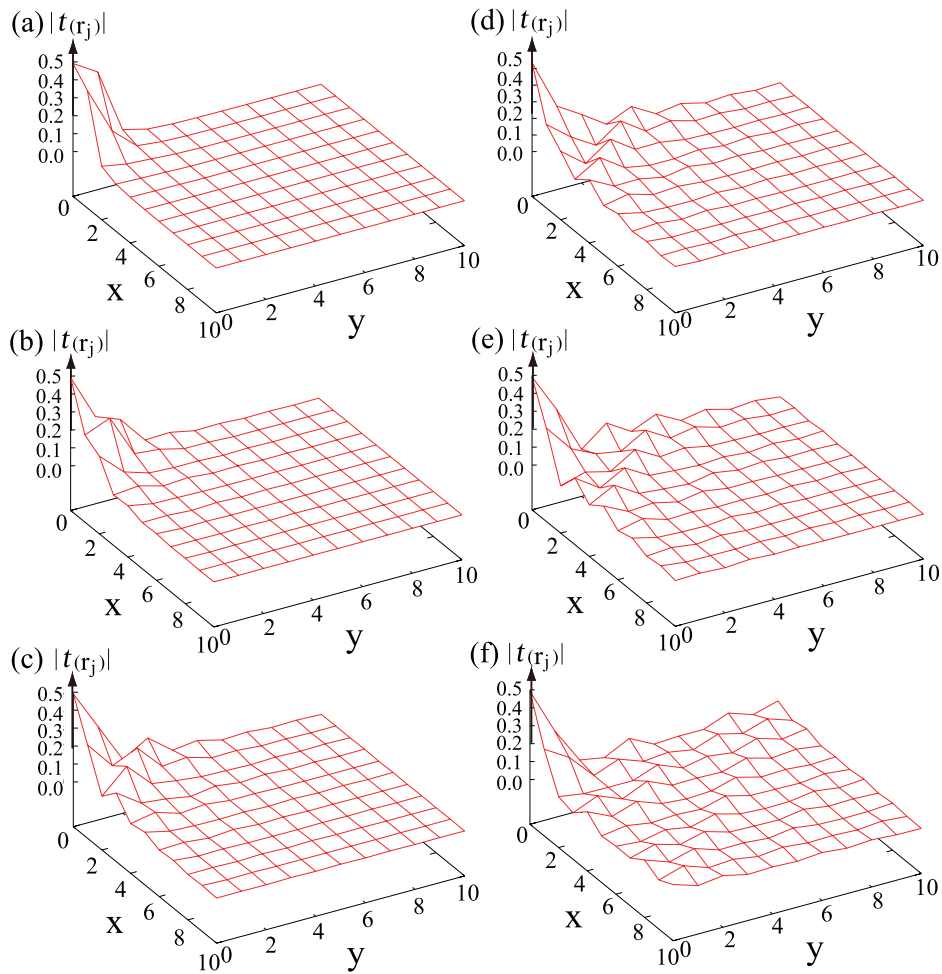}
\end{center}
\caption{\label{Fig0}
The \masa{magnitude} of transfer integrals, \masa{$|{\mathbf t}({\mathbf r}_j)|$} is plotted for the system with $N_x=N_y=100$, for several $C$'s. The values of Chern number are $C=$ (a) 1, (b) 2, (c) 3, (d) 4, (e) 5 and (f) 10. 
}
\end{figure}

In real space, Hamiltonian (\ref{Ham_dn}) can be recast into
\begin{eqnarray}
\mathcal{H}^{(N)}&=&\sum\limits_{{\mathbf R}_j}\sum\limits_{{\mathbf r}_j}c^{\dag}_{{\mathbf R}_j+{\mathbf r}_j\alpha}
\masa{{\mathbf t}^{(N)}({\mathbf r}_j)\cdot\bm{\sigma}_{\alpha\beta}}
c_{{\mathbf R}_j\beta},
\label{Ham_dn_real}
\end{eqnarray}
\masa{where 
\begin{eqnarray}
{\mathbf t}^{(N)}({\mathbf r}_j) = \frac{1}{N_s}\sum_{\mathbf k}{\mathbf d}^{(N)}({\mathbf k})e^{i{\mathbf k}\cdot{\mathbf r}_j},
\end{eqnarray}
with} the number of lattice sites, $N_s$. The summation of ${\mathbf R}_j$ and ${\mathbf r}_j$ are take over the whole lattice.
Indeed, the transfer integrals $t_{\alpha\beta}({\mathbf r}_j)=\bigr({\mathbf t}^{(N)}({\mathbf r}_j)\cdot\bm{\sigma}\bigr)_{\alpha\beta}$ are long-ranged at the cost of complete flatness of energy dispersion. In Fig. \ref{Fig0}
we show the decay of the hopping amplitudes, which is asymptotically exponential with a characteristic range that increases with Chern number.  

\subsection{Effects of truncated hopping range}
\label{finiterange}
The truncation of long-distance components can be carried out by introducing cutoff $d_c$, and
consider only the components satisfying
\begin{eqnarray}
\masa{|n_x|, |n_y|<d_c},\hspace{1cm}{\mathbf r}_j\equiv n_x{\mathbf e}_x + n_y{\mathbf e}_y.
\end{eqnarray}
The resultant truncated Hamiltonian takes the form
\begin{eqnarray}
\mathcal{H}^{(N)}_{\rm trunc}&=\sum\limits_{{\mathbf R}_j}\sum\limits_{{\mathbf r}_j: \masa{|n_x|, |n_y|<d_c}}c^{\dag}_{{\mathbf R}_j+{\mathbf r}_j\alpha}
t_{\alpha\beta}({\mathbf r}_j)
c_{{\mathbf R}_j\beta}\\
&=\sum\limits_{\mathbf k}c^{\dag}_{{\mathbf k}\alpha}{\mathbf d}^{(N)}_{\rm trunc}({\mathbf k})\cdot{\bm\sigma}_{\alpha\beta}c_{{\mathbf k}\beta}
\label{Ham_with_cutoff}
\end{eqnarray}
Here, ${\mathbf d}^{(N)}_{\rm trunc}({\mathbf k})$ contains only the terms like $\cos(n_xk_x+n_yk_y)$ and $\sin(n_xk_x+n_yk_y)$ which satisfy $\masa{|n_x|, |n_y|<d_c}$. The energy dispersions of $\mathcal{H}^{(N)}_{\rm trunc}$ is now given by $\varepsilon=|{\mathbf d}^{(N)}_{\rm trunc}({\mathbf k})|$.
The dispersions are not completely flat, unless $d_c=\infty$. However, if $d_c$ is large enough, (band width)/(energy gap) ratio quickly becomes  very small.
The Chern number also introduces the limitation on the value of $d_c$. From the observation of numerical results, $d_c$ must be $d_c> C$ for Chern number $C$, as we will \masa{discuss} later.

\begin{figure}[t]
\begin{center}
\includegraphics[width=0.79\textwidth]{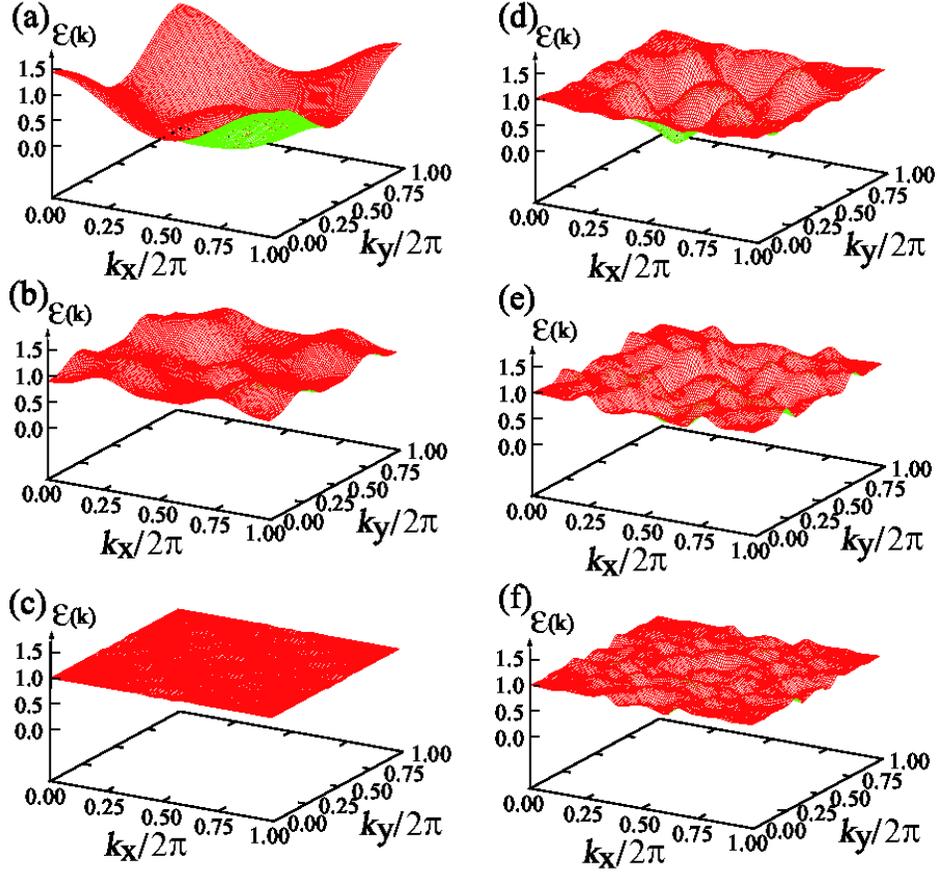}
\end{center}
\caption{\label{Fig1}
The upper band of the two-band Hamiltonian (\ref{Ham_with_cutoff}). (a)-(c) correspond to $C=2$, while (d)-(f) to $C=5$.
The cutoff distances are taken as (a) $d_c=2$,  (b) $d_c=3$,  (c) $d_c=10$,  (d) $d_c=5$,  (e) $d_c=6$,  (f) $d_c=10$.  
}
\end{figure}

We note that the present model is not the shortest range model possible to create a band with Chern number $C$. In fact, the even simpler modification of (\ref{Hofd}), in which we make the replacement \masa{$\mathbf{d}(\mathbf{k})\rightarrow  \mathbf{d}(m_xk_x,m_yk_y)$} 
with any integer \masa{$m_x, m_y$} 
gives a model that has only hopping of range \masa{$m_x, m_y$} 
in the $x$- and $y$-directions respectively, and that has Chern number $C$. Of course, our model (\ref{Ham_dn})  is more "realistic" and homogenous in the sense that its real space representation (\ref{Ham_dn_real}) is dominated by short range terms. Nevertheless, with the alternative construction we see that only a hopping range $\sqrt{C}$ in necessary for obtaining Chern number $C$ in two-band models. However, if the size of the unit cell is allowed to increase, only nearest neighbor hopping is needed \cite{max}. The most intuitive way of creating such models is to couple layers each carrying $C=1$ bands \cite{max,dassarma}. 
Next we consider the energy dispersion of the large $C$ model with a truncated hopping range.
In Fig. \ref{Fig1}, we plot the energy bands for $C=2$ and $5$ with variable cutoff distances, $d_c$.
We show only the energy dispersion $\varepsilon({\mathbf k})$ for upper bands. Due to the particle-hole symmetry, the lower band 
has the symmetric dispersion, $\varepsilon_-({\mathbf k}) = -\varepsilon({\mathbf k})$.

\begin{figure}[t]
\begin{center}
\includegraphics[width=0.7\textwidth]{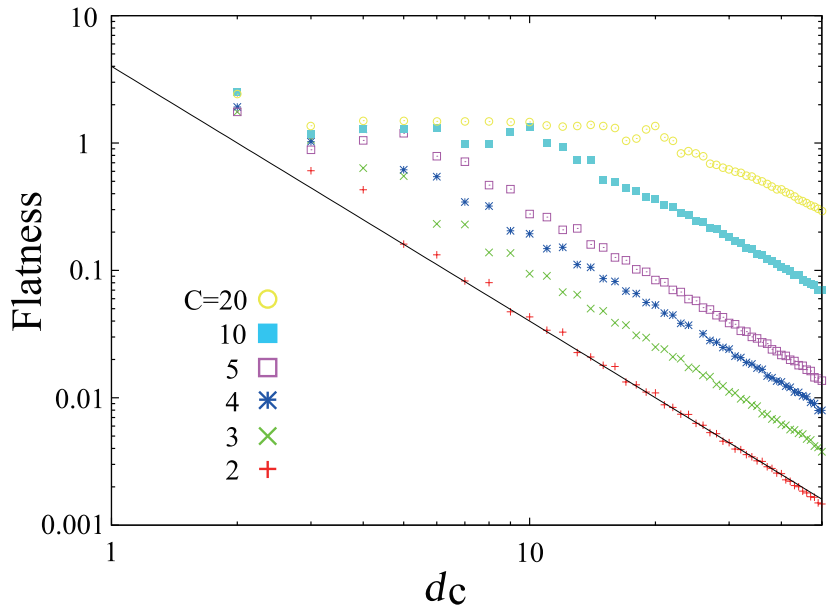}
\end{center}
\caption{\label{Fig_flatness}
$d_c$ dependence of flatness ratio. \masa{The calculation has been done for the systems with $N_x=N_y=100$.} The flatness for $C=2$ is well fitted with $4d_c^{-2}$, as shown with the black line. }
\end{figure}

As shown in Fig.~\ref{Fig1} there is a clear tendency that the band width decreases rapidly, as $d_c$ is increased.
Comparing the band widths between the bands with different Chern numbers, e.g. (c) and (f) of Fig. \ref{Fig1}, we can find larger band width is
obtained for the bands with larger $C$. In Fig. \ref{Fig_flatness}, we plot the flatness ratio of the band, i.e. the ratio of the band with the band gap, as a function of $d_c$.
As noted above, the flatness ratio shows higher value for larger $C$.
As increasing $d_c$, the flatness ratio shows a remarkable drop.
The asymptotic form can be well fitted as $R\propto1/d_c^2$, irrespective of the value of $C$.

While the Hamiltonian (\ref{Ham_dn_real}) leads to Chern number $C$, the introduction of cutoff $d_c$ may affect the Chern number of the band.
In fact, if too small $d_c$ is assumed, the band cannot keep a large Chern number.
To obtain a band with Chern number $C$, $d_c$ has to be set $d_c>C$.
In fact, it is reasonable to require long-range transfer integrals to sustain a band with large $C$.
In terms of the wrapping number interpretation of $C$, it is necessary to include higher harmonics in $\mathcal{H}({\mathbf k})$ to wrap the
sphere surface rapidly enough.
In fact, it requires the $C$-th harmonics, $\cos(Ck_x)$ and $\cos(Ck_y)$, to achieve $C$-fold wrapping while sweeping over the entire Brillouin Zone.
The requirement of long-range hopping also implies a strong non-locality of the electronic state in real space, i.e. local perturbation like on-site impurity potential
affects the electronic property in a broader area for Chern insulator with higher $C$.

\begin{figure}[t]
\begin{center}
\includegraphics[width=0.79\textwidth]{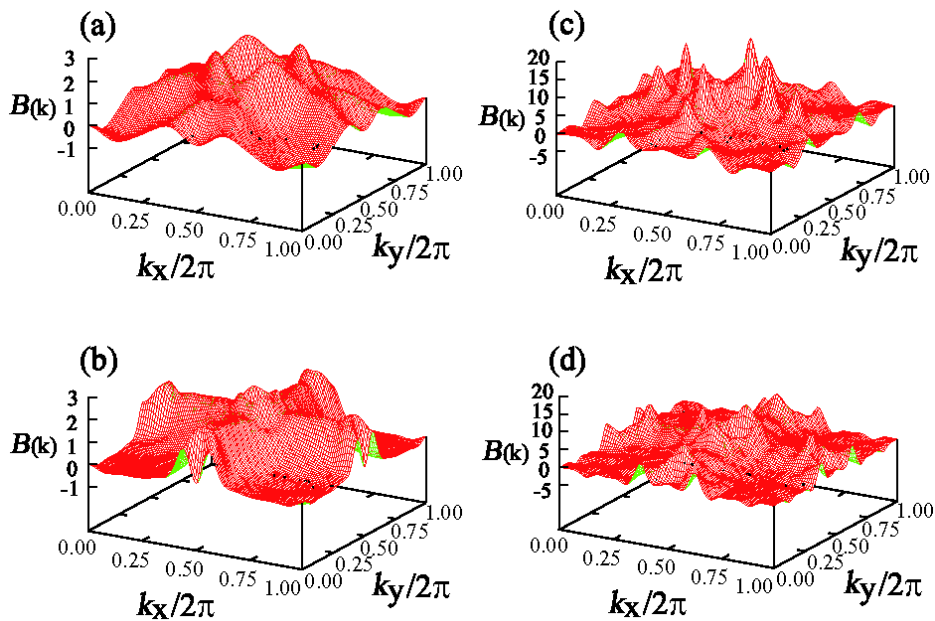}
\end{center}
\caption{\label{Fig2}
Berry curvature of the two-band Hamiltonian (\ref{Ham_with_cutoff}) for (a) $C=2, d_c=3$, (b) $C=2, d_c=10$, (c) $C=5, d_c=6$, (d) $C=5, d_c=10$.   
}
\end{figure}

In Fig. \ref{Fig2}, we plot the Berry curvature for several combinations of $C$ and $d_c$. As shown in Fig. \ref{Fig2}, the Berry curvature shows a strong momentum dependence. As expected from the wrapping number interpretation of the Chern number, Berry curvature changes in a shorter momentum scale for the system with larger $C$. Moreover, in contrast to the energy dispersion, the cutoff $d_c$ does not crucially affect the smoothness of the Berry curvature beyond the critical value needed to create the Chern band in the first place.

This highly dispersive Berry curvature is in sharp contrast to the complete uniformity present for Landau levels.
For the continuum system under a uniform magnetic field, the flat energy dispersion and flat Berry curvature can be simultaneously realized.
It is not clear to what extent a dispersive Berry curvature is harmful to the realization of FCIs (see however Ref. \cite{roy} for a discussion of its possible relevance). In this context, we note that the embedding of a given lattice model in real-space generically changes the Berry curvature distribution (albeit without changing $C$). Nevertheless, if smooth momentum dependence is desired, \masa{interesting approach can be made by resorting to} a relaxation method, by assuming the ${\mathbf d}$-vector to be ``classical spin".
The homogeneous Berry curvature requires uniform distribution of the triple product $\hat{\mathbf d}\cdot\Bigl(\frac{\partial \hat{\mathbf d}}{\partial k_x}\times\frac{\partial \hat{\mathbf d}}{\partial k_y}\Bigr)$ over the BZ.
It may be achieved by assuming an artificial interaction between neighboring ${\mathbf d}$-vector in momentum space, and obtaining its
stable solution by e.g. simulated annealing. Though this procedure cannot achieve completely flat Berry curvature, it may well be an efficient approach towards obtaining a smooth structure.

\section{Entanglement spectrum}
\label{entanglement_formulation}
In this section we consider the one-body entanglement spectrum for bands with higher Chern number.
For this purpose, we cut the system along the $x$ axis, and consider the $N_x\times N_y/2$ strip, which is periodic in $x$ direction,
while open in $y$ direction. The entanglement spectrum can be obtained by Peschel's trick \cite{Peschel,Peschel2}. We introduce the entanglement Hamiltonian,
\begin{eqnarray}
\mathcal{H}_{\rm ES} = \sum_{j_x}\sum_{0\leq j_y<N_y/2}h_{j_xj_y\alpha, j'_xj'_y\beta}a^{\dag}_{j_xj_y\alpha}a_{j'_xj'_y\beta},
\end{eqnarray}
where $a^{\dag}_{j_xj_y\alpha}$ is an auxiliary fermion creation operator defined at site $(j_x, j_y)$ and orbital $\alpha$. The coefficient $h_{j_xj_y\alpha, j'_xj'_y\beta}$ is given by the correlation function obtained from Hamiltonian (\ref{Ham_dn}), assuming the periodic boundary conditions both for $x$ and $y$ directions for the system of the original size: $N_x\times N_y$.

By utilizing the fact that the translational symmetry is still retained in $x$ direction, we make the Fourier transformation to obtain
\begin{eqnarray}
\mathcal{H}_{\rm ES} = \sum_{k_x}\sum_{0\leq j_y<N_y/2}h(k_x)_{j_y\alpha, j'_y\beta}a^{\dag}_{k_xj_y\alpha}a_{k_xj'_y\beta}.
\end{eqnarray}
The matrix $h(k_x)$ can readily be diagonalized to give $N_y$ eigenvalues for each momentum $k_x$, which constitute one-body entanglement spectrum of the system.

\begin{figure}[t]
\begin{center}
\includegraphics[width=0.79\textwidth]{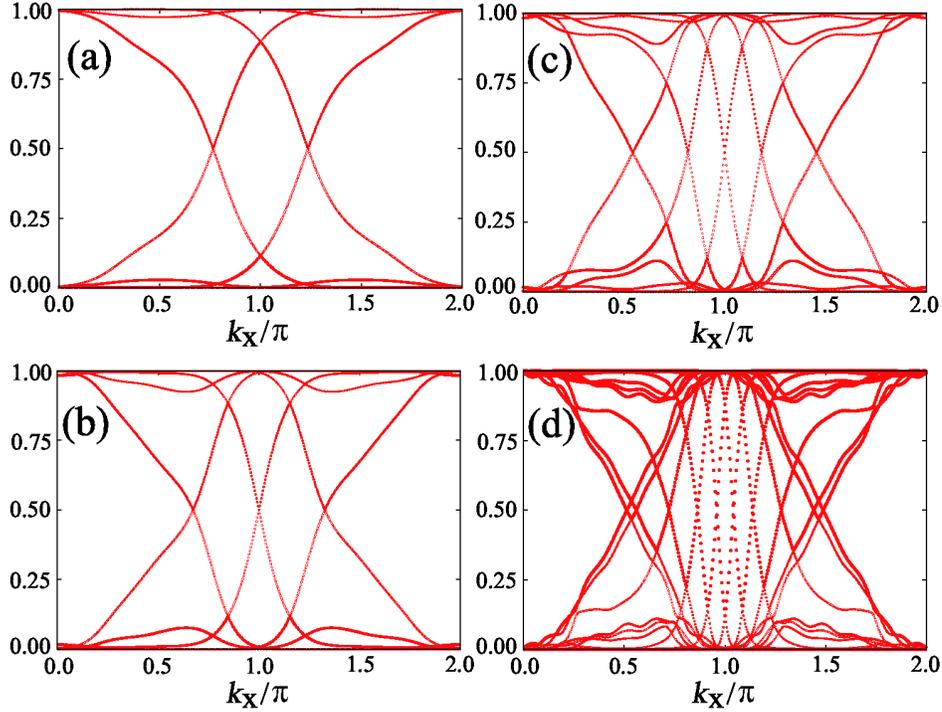}
\end{center}
\caption{\label{Fig2}
Entanglement spectrum for (a) $C=2$, (b) $C=3$, (c) $C=5$, and (d) $C=10$. We set $d_c=2C$. }
\end{figure}

Next, we consider the entanglement spectrum (ES) obtained by bipartitioning the system along the edge.
In Fig.\ref{Fig2}, we plot \masa{part of} the eigenvalues of reduced density matrix against the momenta along the edge.
As clearly shown here, the ES shows $C$-fold crossings for the system with Chern number $C$,
regardless of the details of the system, such as the cutoff $d_c$.

The $C$ modes of entanglement spectrum can be derived from the $C$ fermion edge modes, existing under the open boundary condition  \cite{Turner2010}. 
It may also be related to the strong non-locality in real space for the Chern insulator with large $C$.
Imposing the open boundary condition is equivalent to removing the bonds along the edge from the system with periodic boundary condition.
The existence of $C$ edge modes imply that at least $C$ lanes from the edge are \masa{strongly} affected by the termination of the system.
In other words, a local perturbation affects the state within the distance $C$.
This is closely related to the requirement of long-distance $(> C)$ hopping to sustain the Chern band, which implies the electronic state is 
highly correlated at least within the distance $C$.

\section{Two-body problem}
\label{twobodyproblem}
A key insight, due to Haldane in the context of the FQHE \cite{haldane83}, is that the low energy many-body physics can often be inferred from the much simpler problem of finding the spectrum of few-particle problems. Generalizing the core idea of Ref. \cite{andreas} to systems with $|C|>1$, we consider the energy spectrum of the two-body problem, where the two fermions interact through the short-range interaction:
\begin{eqnarray}
\mathcal{H}_{\rm int} = \frac{\masa{1}}{2}\sum_{\langle i,j\rangle}n_in_j\ \ (n_i=\sum_{\alpha}a_{i\alpha}^{\dag}a_{i\alpha}, \alpha: {\rm orbital\ index})\ .
\label{interaction_Hamiltonian}
\end{eqnarray}
Here, $\langle i,j\rangle$ means summation over the nearest-neighbor sites.
We here consider the situation where the interaction strength is small compared with the band separation, and assume that the
particles are confined to the lower band. Specifically, by introducing the projection operator to the lower band, $P^{(C)}_-$, we consider the following Hamiltonian, 
\begin{eqnarray}
\mathcal{H}_{2}^{(C)} = P^{(C)}_-\mathcal{H}_{\rm int}P^{(C)}_-.
\label{twobody}
\end{eqnarray}
To concentrate on the role of interaction in the projected band, we ignore the kinetic energy part in our analysis on two-body problem.

Suppose that $|\Psi({\mathbf k})\rangle$ is the one-particle eigenstate with momentum ${\mathbf k}$ in the lower band, and $c_{\mathbf k}$ is the annihilation operator for the momentum ${\mathbf k}$ in the lower band, $\mathcal{H}_{2}^{(C)}$ can be modified as follows:
\begin{eqnarray}
\mathcal{H}_{2}^{(C)} &= P_-V\sum_{\langle i,j\rangle}n_in_jP_-\nonumber\\
&\equiv \frac{1}{N}\sum_{{\mathbf k}, {\mathbf Q}, {\mathbf q}} V({\mathbf Q}, {\mathbf k}, {\mathbf q})c^{\dag}_{{\mathbf Q}-{\mathbf k}+{\mathbf q}}c^{\dag}_{{\mathbf k}-{\mathbf q}}c_{{\mathbf k}}c_{{\mathbf Q}-{\mathbf k}}.
\end{eqnarray}
Here,
\begin{eqnarray}
V({\mathbf Q}, {\mathbf k}, {\mathbf q})& = &V(\cos q_x + \cos q_y)\times\nonumber\\ &\times&
\langle\Psi({\mathbf Q}-{\mathbf k}+{\mathbf q})|\Psi({\mathbf Q}-{\mathbf k})\rangle\langle\Psi({\mathbf k}-{\mathbf q})|\Psi({\mathbf k})\rangle.
\end{eqnarray}
The two particle eigenstates can be classified in terms of the center-of-mass momentum, ${\mathbf Q}$ as
\begin{eqnarray}
|\Psi({\mathbf Q})\rangle = \sum_{\mathbf k\in{\rm HBZ}}|\Psi({\mathbf Q}, {\mathbf k})\rangle = \sum_{\mathbf k\in{\rm HBZ}} a({\mathbf Q}, {\mathbf k})c^{\dag}_{{\mathbf Q}-{\mathbf k}}c^{\dag}_{\mathbf k}|0\rangle.
\end{eqnarray}
Here, the summation of ${\mathbf k}$ is restricted to half of Brillouin zone \masa{(HBZ)}, since the wave number ${\mathbf k}'$ and ${\mathbf Q}-{\mathbf k}'$ 
\masa{correspond to the same state}.

The states $|\Psi({\mathbf Q}, {\mathbf k})\rangle$ have the property:
\begin{eqnarray}
|\Psi({\mathbf Q}, {\mathbf Q}-{\mathbf k})\rangle = -|\Psi({\mathbf Q}, {\mathbf k})\rangle
\end{eqnarray}
due to Fermi statistics. By using $|\Psi({\mathbf Q}, {\mathbf k})\rangle$ as basis states, the matrix elements of the Hamiltonian (\ref{twobody}) can be written as
\begin{eqnarray}
\langle\Psi({\mathbf Q}, {\mathbf k}-{\mathbf q})|\mathcal{H}_2|\Psi({\mathbf Q}, {\mathbf k})\rangle = V({\mathbf Q}, {\mathbf k}, {\mathbf q}) - V({\mathbf Q}, {\mathbf k}, 2{\mathbf k}-{\mathbf Q}-{\mathbf q})+\nonumber\\
\hspace{2cm}+ V({\mathbf Q}, {\mathbf Q}-{\mathbf k}, -{\mathbf q}) - V({\mathbf Q}, {\mathbf Q}-{\mathbf k}, {\mathbf Q}-2{\mathbf k}+{\mathbf q}) .
\end{eqnarray}
By diagonalizing this Hamiltonian, one can obtain energies and eigenfunctions for each ${\mathbf Q}$.

\begin{figure}[h]
\begin{center}
\includegraphics[width=0.79\textwidth]{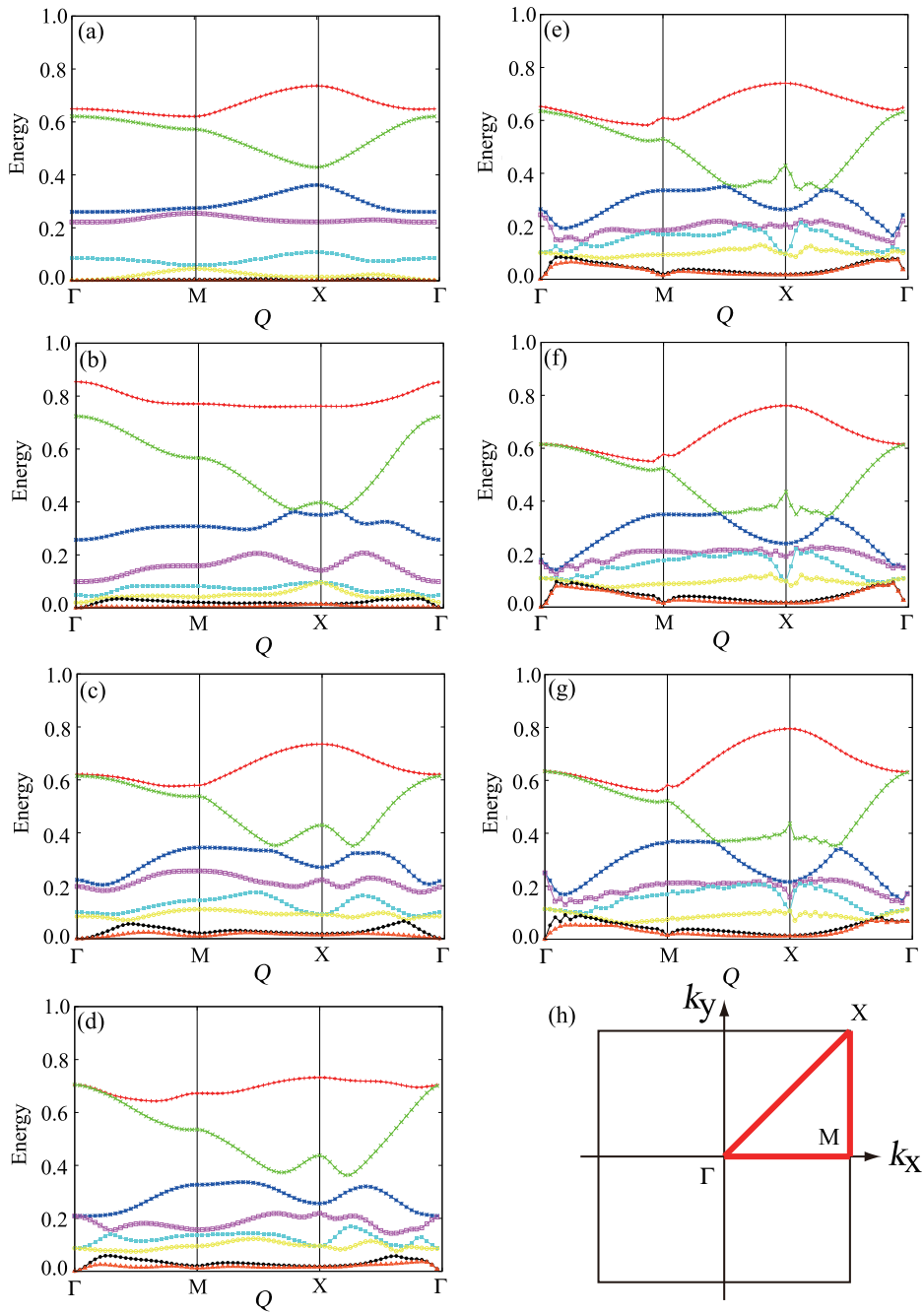}
\end{center}
\caption{\label{Fig.4}
Two-body spectra for (a) $C=1$, (b) $C=2$, (c) $C=3$, (d) $C=4$, (e) $C=10$, (f) $C=15$ and (g) $C=20$ calculated for the system with system size, $N_x=N_y=48$.
We set $d_c=24$.  The spectra are obtained along the symmetry line shown in (h). 
}\label{twobody}
\end{figure}

In Fig. \ref{twobody} we plot the eigenvalues along the representative high symmetry line through the Brillouin zone\masa{, as shown in Fig. \ref{twobody}(h)}.
Notably, the spectrum includes no more than 8 nonzero values for each center of mass momentum. 
In fact it is easy to understand why all other eigenvalues are exactly zero -- before band projection there are precisely 8 non-zero \masa{eigen}values \masa{in Hamiltonian (\ref{interaction_Hamiltonian})}. \masa{In other words, non-zero eigenvalues come from only 8 types of two-particle configurations: two particles each in either of two orbitals ($\alpha$) located next to each other in $x$ or $y$ directions. This means the rank of the Hamiltonian matrix given in (\ref{interaction_Hamiltonian}) is at most 8, and
the rank never increases by multiplying the projection matrices as in eq. (\ref{twobody}).}
In fact, except at 
certain high symmetry points (such as $\Gamma$ in the present case) the possible number of finite eigenvalues will be saturated for a generic local interaction. 

In light of the realization of FQHE, it is favorable to have a well-separated energy spectrum:
By projecting out a part of them, i.e. exactly minimizing the leading terms in the simplified effective problem of independently pairwise interacting particles, a quantum liquid with short-range correlation may be realized, in close analogy with Haldane's pseudo potential argument for the FQHE.
A clear separation of branches is found only for the systems with smaller $C$. This is in most pronounced for $C=1$ (Fig. \ref{twobody}(a)) while for $C=2$ (Fig. \ref{twobody}(b)), some energy gaps may still be inferred for $\varepsilon\sim0.2$ and $0.8$. A closer analysis however, reveals that only gaps between pairs of energy levels are leading to stable FCIs \cite{andreas,zhaoeliot}, thus already for $C=2$, it is likely that the present model does not support stable FCI states. Increasing $C$, the gaps close entirely (Fig. \ref{twobody}(c)-(g)), and the spectrum develops highly entangled features. 

This is a further indication that the present choice of tight-binding model in combination with the specific interaction is not particularly favorable for realizing FCIs. Thinking more concretely about the variable $C$ bands in terms of $C$ coupled $C=1$ bands it is indeed likely that a separation of energy scales, or at least a clear gap in the two-particle spectrum, would be needed to achieve FCI ground states. More specifically, if the largest two eigenvalues would be much larger than the others we would expect the realization of FCIs at filling fractions $\nu=\frac 1{2C+1}$ \cite{jorg}. 
Moreover, in $C=1$ models it has been shown that "interaction flatness" is a good indicator for the suability of a given model for realizing FCIs \cite{zhaoeliot}. Also this criteria is increasingly violated for higher $C$. Taken together, these observations provide a rationale for the apparent lack of realizations of FCIs in flat bands with high Chern numbers in two-band models. In fact, it was only once flat band models with unit-cells increasing in size with $C$ \cite{max} that FCIs were systematically identified \cite{ChernN,ChernN2}.

\section{Discussion}\label{discussion}

In this work we have approach\masa{ed} several basic questions concerning flat band models with varying Chern numbers through explicit examples mainly using a simple class of two-band models. To wrap up we will now make a few remarks on more generic models and relate to other results in the literature while summarizing our main conclusions. 

It is known that any two of the features strictly finite hopping range, non-zero Chern number and exactly flat dispersion are realizable simultaneously, while all three criteria cannot be met simultaneously \cite{nonflat} as long as the number of bands remains finite. It is also known that and exponential tail of hopping processes is enough to flatten the topological $C\neq 0$ bands \cite{kapit}.  
In this work, these facts were explicitly illustrated using our simple two-band model, for which we also found that the minimal hopping range needed  to achieve Chern number $C$ is $d_c> C$. We note however that this is not optimal, and that one can easily write down two-band models with hopping range $\sqrt{C}$ supporting (non-flat) bands with Chern numbers $\pm C$. By merging independent $N_{{\rm bands}}/2$ copies of such models directly leads to models with a hopping range of $\sqrt{4|C| /N_{{\rm bands}}}$ and bands with Chern numbers $\pm C$. 
Interestingly, the scaling of this very simple construction coincides (up to a constant factor) with that of the related problem of finding the minimal mean hopping range while requiring entirely flat bands, which was argued to be $\sqrt{4|C|/\pi N_{{\rm bands}}}$ by the authors of Ref.~\cite{instanton}.

The Berry curvature fluctuates regardless of hoping range, thus responding in a fundamentally different way to changes in the hopping amplitudes as compared to the energy dispersion. In fact, the original Dirac models (\ref{Hamiltonian},\ref{HofK}) have precisely the same eigenstates as the corresponding (spectrally) flattened models including long-range hopping, hence their Berry curvature is identical throughout the Brillouin zone. Consistent with this, we found that truncating the hopping range by a finite $d_c$ in our flattened models only mildly changes the Berry curvature distribution as long as $d_c> C$. We note that, by explicit construction, one can find models for which the Berry curvature variations decrease exponentially as a function of the total number of bands and in the continuum limit (cf. Landau levels) the Berry curvature is entirely flat. To the best of our knowledge, it remains however as an open problem to find fundamental lower limits on the Berry curvature variations as a function of  Chern number and the total number of bands.

We also investigated the entanglement spectrum \cite{LiH} of our model. Although recent works have stressed the potential pitfalls with inferring universal information from numerically obtained entanglement spectra \cite{Sondhi,Budich}, the correspondence between the (physical) edge theory and the bulk entanglement \cite{LiH,Peschel2,Turner2010} worked out rather beautifully in the present case. Indeed, we generically found $|C|$ branches of each chirality transversing the entanglement gap, precisely analogously with the $|C|$ chiral edge state channels crossing the energy gap at each edge when the physical system is put on a geometry with open boundaries (in this case a cylinder). We also found that these results were very robust to changes in the hopping amplitudes as long as the Chern number remained unchanged, as one would indeed hope for. 

Finally, we saw that the two-particle spectrum of a local interaction projected to a Chern band quickly became jagged and lacking a separation of energy scales when $C$ is increased. This scenario is  unfavorable for stabilizing FCIs, and provides a heuristic understanding of why FCIs are increasingly likely to occur in models with more bands. In retrospect, these observations provide a rationale for why the realization of FCIs in flat bands with increasingly large $C$ \cite{ChernN,ChernN2} was first made possible by the introduction of flat band models with correspondingly larger unit-cells \cite{max,dassarma}.

Taken together, our observations reported in this work are likely to prove useful in future studies of the interacting phase diagram of flat band models with various Chern numbers, which still remain largely unexplored beyond the Landau-level like case of $C=1$.

\ack
We thank J\"org Behrmann, Jan Budich, Jens Eisert and Roderich Moessner for related discussions.  E.~J.~B. acknowledges support from DFG's Emmy Noether program (BE 5233/1-1). M. U. is supported by Grants-in-Aid for Scientific Research (No. 24340076, 26400339, and 24740221), Japan.

\end{document}